# Soliton crystal Kerr microcombs for high-speed, scalable optical neural networks at 10 GigaOPs/s

Xingyuan Xu*, Mengxi Tan, David J. Moss

*Optical Sciences Centre, Swinburne University of Technology, Hawthorn, VIC 3122, Australia*
*\*Current Address: Dept. of Electrical and Computer Systems Engineering, Monash University, Clayton, 3800 VIC, Australia*

*Abstract*— Optical artificial neural networks (ONNs) have significant potential for ultra-high computing speed and energy efficiency. We report a new approach to ONNs based on integrated Kerr micro-combs that is programmable, highly scalable and capable of reaching ultra-high speeds, demonstrating the building block of the ONN — a single neuron perceptron — by mapping synapses onto 49 wavelengths to achieve a single-unit throughput of 11.9 Giga-OPS at 8 bits per OP, or 95.2 Gbps. We test the perceptron on handwritten-digit recognition and cancer-cell detection — achieving over 90% and 85% accuracy, respectively. By scaling the perceptron to a deep learning network using off-the-shelf telecom technology we can achieve high throughput operation for matrix multiplication for real-time massive data processing.

## I. INTRODUCTION

Artificial Neural Networks have achieved impressive success at extracting predictions and simple representations from complex high dimension data. When adequate data is used for training, ANNs can outperform computational algorithms [1-5] and even humans for many tasks ranging from the translation of languages to recognition of images, evaluation of risks and, interestingly enough, even complex board games [6]. The computational power and speed of ANNs is governed by matrix multiplication operations. Current electronic chips for ANNs include the Google TPU and IBM TrueNorth chips [7, 8]. They use extremely large scale processor arrays that include the systolic array [8], to enhance the parallelism to achieve operational speeds exceeding 180 trillion floating point operations a second (Tera-FLOPS). Despite this, though, they are still subject to relatively inefficient digital protocols or bandwidth bottlenecks due to the electronics, or both – each single processor is limited in speed to only about 700 MHz [9].

Optical neural networks (ONNs) are photonic approaches towards ANNs that are new systems geared towards the next generation of neuromorphic processing. They are highly promising since they offer the potential to achieve extremely high processing speeds [3]. The critical issue is to be able to achieve the weighted synapses that connect the nodes and neurons. As opposed to electronic digital systems that store the synapses in memory, photonic based methods operate through the actual physical implementation of synapses, where their number determines the scale of the network, and which depends on the physical parallelism that is fundamentally analog in nature.

Substantial success has been achieved for ONNs, and these have relied on a variety of strategies to multiplex many synapses in parallel. Schemes that rely on spatial multiplexing include diffractive optics [10] as well as coherent integrated photonic chips [3]. These have been successful at achieving tasks for classifying alphabetic characters (vowels) or handwritten numbers (digits). Furthermore, they have demonstrated operation at very low power levels, albeit they are subject to a compromise between the physical system size in terms of footprint, and processing power of the system, determined by the number of synapses or degree of parallelism. There are a variety of strategies to implement ONNs, and these include photonic reservoir computing [11-13] as well as a method called spike processing [14-17]. These both use sophisticated schemes to multiplex the synapses and are both very compact. Reservoir computing with photonics multiplexes the synapses temporally in order to realise systems on a very large scale, and with 100's of nodes at the input layer. On the other hand, spike processors, that have successfully demonstrated pattern recognition tasks by exploiting phase change based integrated devices [17], operate via wavelength division multiplexing (WDM). This approach also benefits from having an operation bandwidth that is dynamically reconfigurable [10-12]. Despite their success, these approaches still face limits of various kinds. Temporal multiplexing is challenging to both train in a dynamic way, or to scale up to achieve deep learning systems that comprise multiple layers. Spike processing is restricted in the amount of parallelism that it can achieve because of its reliance on arrays of discrete laser diodes. The combined use of temporal, wavelength, and spatial multiplexing has the greatest potential to achieve the highest processing power, operation speed and scale of the network, and this is what our system is based on.

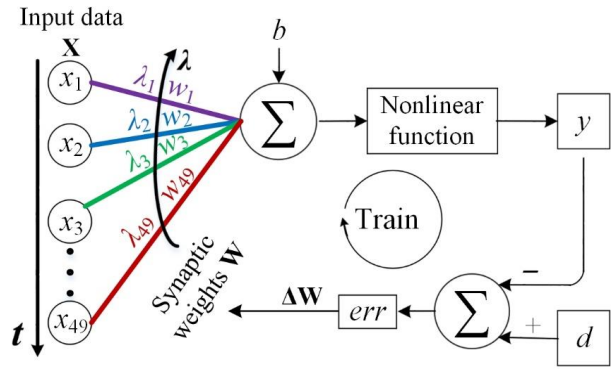

Figure 1. Mathematical model of perceptron.

## II. PERCEPTRON

In this paper, we report [4] a novel scheme for ONNs that relies on integrated micro-combs to achieve simultaneous multiplexing in the temporal, spatial, and wavelength domains at the same time, in order to calculate the dot product of vectors. We perform matrix operations by flattening the vectors first to convert them into matrices at high data rates. Our system is capable of dynamic training and its network structure is highly scalable. We demonstrate a single photonic neuron perceptron that has 49 synapses. This fundamental component of ONNs achieves a speed for matrix multiplication of 11.9 billion (Giga) operations/s (OPS) – or GOPs - that equates to 95.2 Gigabits/s for 8 bit operations. We do this via simultaneous synapse weighting in the wavelength domain and the temporal domain, scaling the input data. The device is applied to benchmark tests that include handwritten digit classification, where we obtain an accuracy greater than 93%, and to the prediction of cancer classes to distinguish malignant from benign cases based on an extracted feature set from microscope images from biopsied tissue. We obtain an accuracy of greater than 85% for the cancer classification.

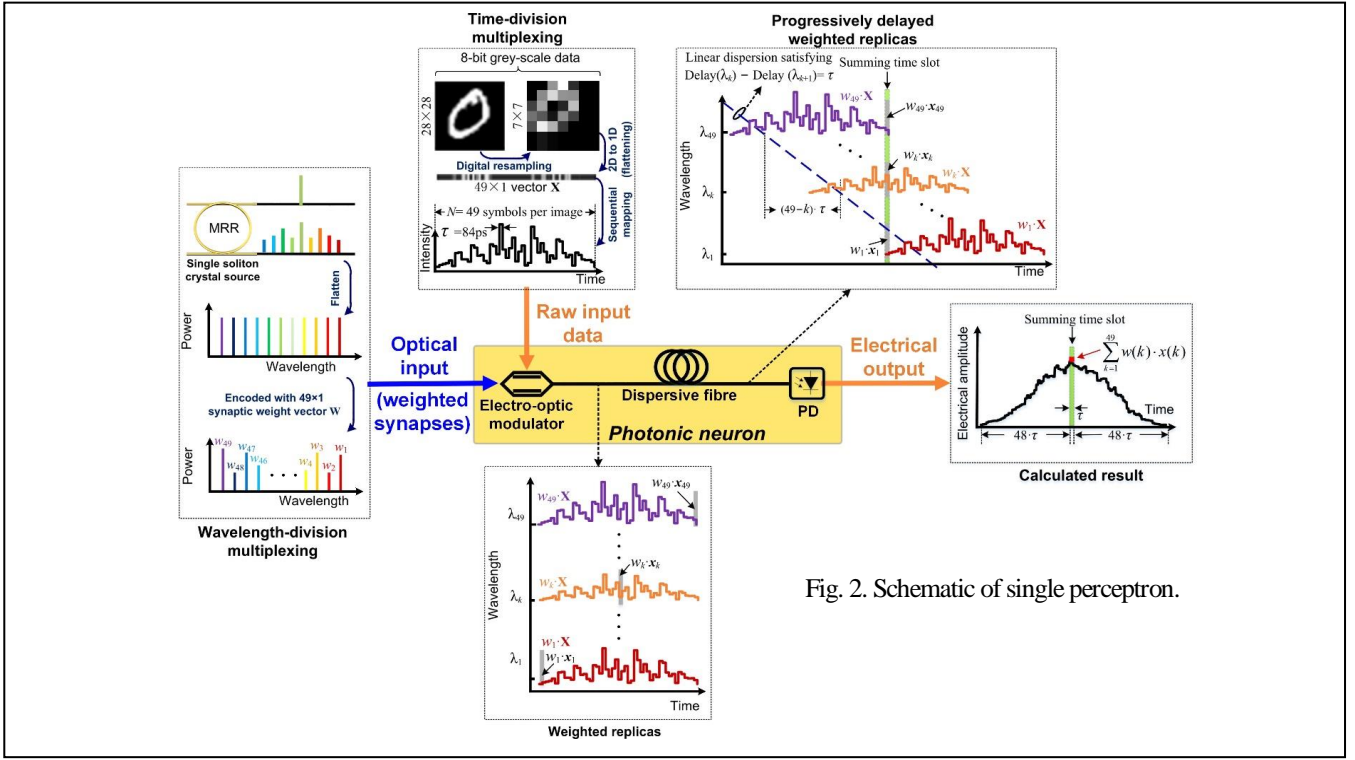

Fig. 2. Schematic of single perceptron.

Figure 1 depicts the neuron perceptron mathematical model [18] and Figure 2 outlines the experiment setup that uses a Kerr optical micro-comb source. The perceptron is based on wavelength multiplexing with 49 microcomb wavelengths, simultaneously with temporal multiplexing, to form a single synapse. The main operation consists of matrix multiplication with vectors formed from flattened matrices. The matrix multiplication occurs between the electronic image input data and the synaptic weights, and this is performed in a multi-step way in the photonic scheme. The input data for classification consists of $28 \times 28$ electronic digital matrices with 8-bit grey-scale intensity resolution, which is initially down sampled digitally into 7×7 matrices that are then reorganized into one dimensional vectors: $X(i) = [X(1), X(2) … X(49)]$, which are then multiplexed sequentially in the temporal domain by an electronic high speed D/A converter at 11.9 Gigabaud. Here, each symbol corresponds to the 8-bit pixel input data images and takes up one time slot 84 ps in length. Hence, the whole duration of the waveform is N x τ = 4.12 ns with N=49. In conventional approaches based on digital electronics, the neural network input nodes usually reside in electrical memory and are routed according to memory address. By comparison, the input nodes for our ONN are temporally defined by multiplexing the symbols that are then routed, determined by their location in time.

Following this, the electrical input waveform that is a temporally multiplexed signal is broadcast via an electro-optic modulator on to all 49 wavelengths (equal to the number of elements of the vector X), the wavelengths being generated by the micro-comb. Here, each comb line contains an equal copy of X, the time domain multiplexed input data waveform. Every comb line's power is then adjusted by an optical waveshaper with the weights being determined by the theoretical synaptic weight vector $W = [w(1), w(2), …, w(49)]$ obtained during training. This then, in effect, serves to multiplex the synaptic weights in wavelength. If W and X are both 1×49 column vectors, then the weighted input X vector replicas can be written as

$$\mathbf{X} \times \mathbf{W}^T = \begin{pmatrix} w(1) \cdot x(1) & w(1) \cdot x(2) & w(1) \cdot x(3) & \cdots & w(1) \cdot x(49) \\ w(2) \cdot x(1) & w(2) \cdot x(2) & w(2) \cdot x(3) & \cdots & w(2) \cdot x(49) \\ w(3) \cdot x(1) & w(3) \cdot x(2) & w(3) \cdot x(3) & \cdots & w(3) \cdot x(49) \\ \vdots & \vdots & \vdots & \ddots & \vdots \\ w(49) \cdot x(1) & w(49) \cdot x(2) & w(49) \cdot x(3) & \cdots & w(49) \cdot x(49) \end{pmatrix}$$

(1)

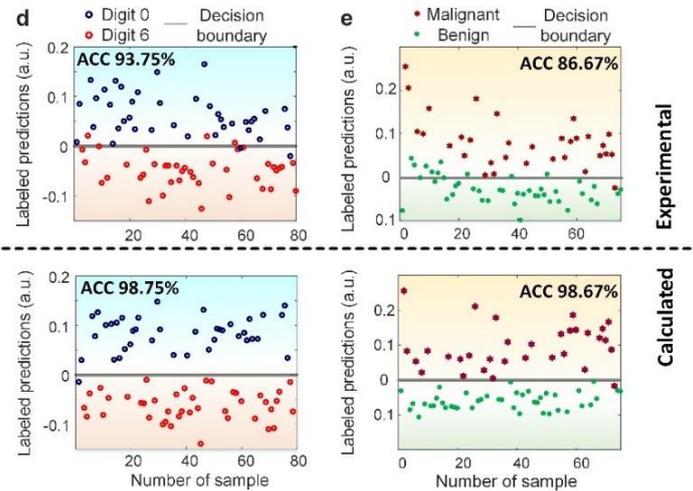

Fig. 3. Results. d, Predictions for handwritten digit recognition showing an accuracy of 93.75% vs 98.75% for theory. e, Predictions for benign versus malignant tumor cells with an accuracy of 86.67% versus 98.67% for theory.

where the nth row (where $n \in (1, N)$) corresponds to the temporal weighted waveform replica of the $n^{th}$ wavelength. Therefore, the diagonal components reflect the input N weighted nodes, so that the $n^{th}$ weighted input node is reflected in the 8-bit symbol $w(n) \cdot x(n)$ that exists in the $n^{th}$ time slot for the $n^{th}$ wavelength. After this, the replicas are transmitted through a medium that provides a dispersive delay equivalent to $2^{nd}$ order dispersion, to sequentially delay the weighted replicas in order to align the diagonal components into the same time window, with the step in delay being given by $\tau = \text{delay}(\lambda_k) - \text{delay}(\lambda_{k+1})$. Therefore, the dispersive delay component acts as an addressable time-of-flight memory that lines up the progressively weighted time dependent symbols w(1) · x(1), w(2) · x(2) ... w(49) · x(49) over all wavelengths as

$$\begin{pmatrix} & & & w(1)\cdot x(1) & \cdots & w(1)\cdot x(47) & w(1)\cdot x(48) & w(1)\cdot x(49) \\ & & w(2)\cdot x(1) & w(2)\cdot x(2) & \cdots & & w(2)\cdot x(48) & w(2)\cdot x(49) \\ & w(3)\cdot x(1) & w(3)\cdot x(2) & w(3)\cdot x(3) & \cdots & & & w(3)\cdot x(49) \\ \cdot\cdot\cdot & \vdots & \vdots & \vdots & \cdot\cdot\cdot & & & \\ w(49)\cdot x(1) & \cdots & w(49)\cdot x(47) & w(49)\cdot x(48) & w(49)\cdot x(49) & & & \end{pmatrix}$$

While this process does not enhance the network speed because it only uses diagonal components, in principle a significant increase in speed can be obtained by scaling the network to deep (multiple level) structures through the use of parallel wavelengths as well as time and spatial multiplexing.

Finally, the intensity of all of the optical signals that are in the time bin that is lined up, are all summed via sampling and detection to produce the resulting matrix multiplication (equivalent to a dot product of 49×1 vectors for the case of 7×7 matrices) of the neuron, given by:

$$\mathbf{X} \cdot \mathbf{W} = \sum_{k=1}^{49} w(k) \cdot x(k) \quad (3)$$

After matrix multiplication, the summed, weighted output is then modulated in order to map it into a desired range by using a nonlinear sigmoid function. In this initial demonstration we achieve this last function by digital electronics, which yields the output of the single neuron perceptron. In principle, however, this can readily be achieved all-optically. Finally, the input data prediction category is produced through comparison between the decision boundary with the neuron output. The decision boundary is a hyperplane in a space with 49 dimensions, which is generated during digital learning carried out beforehand offline. Thus, the input data can be separated into the two categories.

### III. MICRO-COMBS BASED ON SOLITON CRYSTALS

Kerr optical micro-combs [19-21] have formed the basis for a multitude of successful breakthroughs including optical frequency synthesis [22], ultra-high bit rate data transmission [23], generation of advanced quantum states [24], high level RF signal processing [25], and much more. They provide the full capability of mainframe optical frequency combs [26] although in a fully integrated form that has a much more compact footprint as well as the potential to scale the network in power, reliability, and performance [27-40]. The new platforms developed for optical microcombs have much lower nonlinear absorption than other nonlinear platforms such as chalcogenide glass and semiconductors [41-61].

We use a microcomb that operates via soliton crystal states, produced in integrated ring resonators. Soliton crystals display deterministic generation induced by mode crossings that produce a background wave, with all of these processes sustained by the Kerr nonlinearity. For soliton crystals, there is very little nonlinear pump-induced shift in the resonance that otherwise would require difficult dynamic pumping schemes like DKS solitons require [27]. This is because the intracavity soliton crystal state power is virtually the same as the power for the chaotic state from which it is formed. Hence, very little power jump occurs when they are generated and this allows a reliable and simple method of initiation achievable by simple adiabatic, even manual, tuning of the pump wavelength [62]. This same effect also yields a much higher energy conversion efficiency from pump to comb-line [63]. Soliton crystals have demonstrated a multitude of RF or microwave signal processing based on photonics [25, 64-88]. The integrated ring resonators were made from Hydex glass, a platform that is CMOS compatible [20]. They had a high Q factor of 1.5 million and a 48.9 GHz FSR with a chip to fibre coupling loss of 0.5 dB / facet achieved by on-chip mode converters. The waveguide cross-section of 3µm × 2µm produced anomalous dispersion with a mode crossing near 1552 nm. A 30dBm CW pump laser generated the soliton crystals when its wavelength was swept manually from short to long wavelengths (blue to red) near a resonance.

### IV. RESULTS

First, we evaluated the performance of the network using a number of handwritten digit pairs from a body of 500 images for each digit, from which we randomly selected 920 images for prior off-line training, which left 80 figures to evaluate the system performance. The handwritten digital images were electronically down-sampled to reduce the size of the images to 7×7 from 28×28. Next, this was transformed into a 49 symbol one dimensional array, following which the array was temporally multiplexed with each symbol occupying an 84ps time slot, yielding a modulation rate of 11.9 Gigabaud. The data vector dimension of our perceptron needed to match the weight vector dimension, given by number of wavelength, which was 49. Therefore, we used a down-sampling method on the image to reduce the length of the vector to 49.

The optical power for each of the 49 comb lines was weighted according to the pre-learned synaptic weights in order to enhance the parallelism to form the neuron synapses. Next, the data input stream was simultaneously imprinted onto all of the 49 weighted microcomb lines, which were then linearly progressively delayed in wavelength by 13km of single mode fibre that generated a time-of-flight optical buffer via its 2$^{nd}$ order dispersion of 17 ps / nm / km. Therefore, the weighted symbols for each wavelength were aligned in time, thereby enabling them to be summed by simple sampling of the centre timeslot and subsequent detection. This yielded the matrix multiplication result, a product of the multiply and accumulate (MAC) operation. The output was finally compared against the decision boundary which consists of a hyper-plane that was generated during prior network training that classified the input samples arranged in a 49-dimensional hyperspace. The resulting matrix multiplication computations on the multiple input data samples were then compared in intensity against this decision boundary, finally producing the predictions of the perceptron (Fig. 3). We tested the perceptron performance for classifying 2 benchmark tests delineated by the decision boundary – first for two handwritten digits (0 and 6), followed by determining whether cancer cells are benign or malignant. For the handwritten digits the perceptron produced an accuracy of 93.75%, versus 98.75% that can be achieved with an electronic digital neural network. For the tissue biopsy data classification for cancer cells (Fig. 3), individual cell nuclei were extracted from breast mass tissue via fine needle aspirate and then imaged with a microscope. These images were previously characterized to distinguish 30 different features including texture, perimeter, radius, etc.. For our experiments, the data for 521 cell nuclei were used for pre-training the network, with a further 75 used as the basis for the testing diagnosis. This follows a very similar process to that used for the handwritten digit tests discussed above. We obtained an 86.67% accuracy versus 98.67% that can be achieved with a digital electronic neural network.

In our experiments we used Intel's approach of evaluating digital microprocessors [89]. Since our system is rather more complex in that it uses input data and weight vectors for the MAC calculations that come from different sources that are multiplexed in time and wavelength, we define the throughput speed based on the temporal data sequence of the electronic output port, in order to be unambiguous. According to the protocol of broadcast-and-delay, each computation cycle consists of one vector dot product between the 49 symbol data and the weighted vectors, resulting in a time data sequence having a length of 48+1+48 symbols, yielding a total duration time of 97 × 84ps. The 49$^{th}$ symbol represents the desired result – ie., the vector dot product resulting from 49 MAC operations, and hence the perceptron throughput is given by 49 / (84ps × 97) = 5.95 Giga-MACs/s. Since each MAC operation consists of two operations — a multiply followed by an accumulate operation —our throughput measured in operations (OPS) is twice that measured in MACs/s, or (49×2)/(84 ps×97) = 11.9 Giga-OPS.

The input data sequence contained 8-bit symbols of 256 discrete levels, reflecting the pixel values of the grey scale image. The 8 bits was limited by our electronic arbitrary waveform generator's intensity resolution. The Waveshaper had a range in attenuation of 35 dB, which is equivalent to a resolution of 11 bits or 33 dB ($=10\times\log_{10}[2^{11}]$). Therefore, every computing cycle had an effective throughput bit rate of $(49\times2) \times 8 / (84\text{ ps} \times 97) = 95.2$ Gigabits/s. For analogue systems such as ours, both the intensity resolution and the bit rate are limited by the system SNR (signal-to-noise ratio). Therefore, in order to have a full resolution of 8-bits, our system needed to have a SNR greater than $20\cdot\log_{10}(2^8) = 48$ dB in terms of electric power. This is well within the capability of analogue photonic microwave links, such as the perceptron system that we reported here which had an OSNR >28 dB.

Our perceptron is the fastest optically based neuromorphic processor ever reported, although making direct comparisons with all of the different approaches is challenging since they vary so widely. As an example, on the one hand systems based on static or continuous sources that perform one-off or single-shot measurements [4, 10, 17] can have a very low latency. However, on the other hand, they also suffer from an extremely low throughput since the input data cannot be in any rapid manner. While our perceptron did have a relatively large latency of ~64 μs, this was purely due to the dispersive delay component which in our case was a simple spool of optical fibre. This did not, however, have any effect on the speed or throughput of our system. Moreover, in fact this can be dramatically reduced or virtually eliminated – easily to < 200 ps – just by using any type of compact device that can replace the dispersive delay of the fibre, such as sampled Bragg gratings or etalon based tuneable dispersion compensators [90-95]. Finally, this approach of temporal, wavelength and spatial multiplexing can be applied to a wide range of neuromorphic processors including convolutional accelerators and deep learning convolutional neural networks [96].

## V. Conclusions

We report an optical neural network consisting of a single perceptron that operates through the use of an integrated optical Kerr micro-comb source. The system achieves a single processor throughput speed of 11.9 Giga-OPS/s, equivalent to 95.2 Gigabits/s. We demonstrate standard real-life benchmark tests including cancer cells diagnosis and handwritten digit recognition. We outline different approaches in order to expand the network to deep learning ONN architectures that have significantly increased processing power and throughput speed. This is possible because of the high level of parallelism that can be realized via simultaneous time, spatial, and wavelength multiplexing. Our approach has enormous possibilities for the real-time analysis of very high dimensional data that demanding applications will need.